# Critical decay index for eruptions of 'short' filaments

Boris Filippov

Pushkov Institute of Terrestrial Magnetism, Ionosphere and Radio Wave Propagation of the Russian Academy of Sciences (IZMIRAN), Troitsk, Moscow 108840, Russia
(e-mail: bfilip@izmiran.ru)

**Abstract**

Model of a partial current-carrying torus loop anchored to the photosphere is analyzed. Conditions of the catastrophic loss of equilibrium are considered and corresponding value of the critical decay index of external magnetic field is found. Taking into account line-tying conditions leads to non-monotonous dependence of the critical decay index on the height of the apex and length of the flux rope (its endpoints separation). For relatively short flux ropes, the critical decay index is significantly lower than unity, which is in contrast to widespread models with the typical critical decay index above unity. The steep decrease of the critical index with height at low heights is due to the sharp increase of the curvature of the flux-rope axis that transforms from a nearly straight line to a crescent.

**Key words**  Sun: activity - Sun: filaments, prominences – Sun: magnetic fields.

## 1 Introduction

Magnetic flux ropes in the solar corona are considered as the most probable magnetic configurations in source regions of eruptive prominences and coronal mass ejections (CMEs). Coronal cavities and solar filaments (prominences) serve as tracers of flux ropes in coronal images. Stretching of these structures along photospheric magnetic polarity inversion lines (PILs) indicates the presence of non-neutralized electric currents within the coronal flux ropes, which 'feel' the external magnetic field due to the action of the Lorentz force. A PIL in the corona is a place where a coronal electric current (a flux rope) can be set in horizontal equilibrium. The height of the equilibrium position depends on the strength of the electric current as well as of the coronal magnetic field (van Tend & Kuperus 1978; Molodenskii & Filippov 1987; Demoulin & Priest 1988; Martens & Kuin 1989; Priest & Forbes 1990; Forbes & Isenberg 1991; Lin et al. 1998; Schmieder et al. 2013). However, the stable equilibrium is possible only if the background field decreases with height not too fast, or the decay index of the ambient magnetic field does not exceed a critical value $n_c$. (van Tend & Kuperus 1978; Bateman 1978; Forbes & Isenberg 1991; Lin et al. 1998; Kliem & Torok 2006; Fan & Gibson 2007; Isenberg & Forbes 2007; Demoulin & Aulanier 2010; Olmedo & Zhang 2010; Nindos et al. 2012).

The exact value of $n_c$ depends on parameters of a flux-rope model. For a thin straight current channel $n_c = 1$ (van Tend & Kuperus 1978; Filippov & Den 2000, 2001). For a thin circular current channel $n_c = 1.5$ (Osovets 1961; Bateman 1978; Kliem & Torok 2006). Kliem & Torok (2006) called instability of a toroidal current ring immersed in background potential field 'torus instability'. Demoulin & Aulanier (2010) showed that the critical decay index $n_c$ has similar values for both the circular and straight current channels in the range 1.1 - 1.3, if a current channel expands during an upward perturbation, and in the range 1.2 - 1.5, if a current channel would not expand. Numerical MHD simulations of relaxations of nonlinear force-free equilibria suggest values of the critical decay index in the range $n_c = 1.4 - 1.9$ (Torok & Kliem 2005, 2007;

Fan & Gibson 2007; Isenberg & Forbes 2007; Aulanier et al. 2010; Fan 2010; Kliem et al. 2013; Amari et al. 2014; Inoue et al. 2015; Zuccarello et al. 2015, 2016).

The decay index was computed in regions above the photosphere from a potential field extrapolation (Liu 2008; Guo et al. 2010; Kumar et al. 2012; Nindos et al. 2012; Xu et al. 2012) or a nonlinear force-free field extrapolation (Liu et al. 2010; Cheng et al. 2011; Savcheva et al. 2012) in order to find the difference in coronal magnetic fields for failed eruptions and full eruptions, a threshold of flux-rope instability, the relationship between the decay index and CME speed, and so on.

Zuccarello et al. (2014) in the study the evolution of an active region filament and Su et al. (2015) in the study the eruption of a polar crown prominence found that at the moment of the eruption the critical decay index was $n_c \approx 1$. McCauley et al. (2015) studied the kinematics of more than one hundred limb prominence eruptions and found that the decay index at the onset of the fast-rise is in the range 0.7–2 with a mean value of $n_c \approx 1.1$. The average decay index for eruptive filaments in 2012 -- 2013 in the study of Aggarwal et al. (2018) has fairly broad distributions in the range 0 -- 2, with a mean value of 0.84. Sarkar et al. (2019) found that before the eruption on 2010 June 13 the top of a polar crown prominence reached the height with the decay index value of 1.2, while the decay-index value at the corresponding cavity centroid height reached the critical limit of 1.4 for the onset of torus instability. In the study of 16 failed filament eruptions observed near 24 solar cycle maximum, there were found that more than half of studied events begun in the region with the decay index value lower than unity (Filippov 2020).

In this paper, we consider a model of a partial current-carrying torus loop anchored to the photosphere. We analyze conditions of the catastrophic loss of equilibrium and find the dependence of the critical decay index on length of the flux rope (its endpoints separation) relative the height of the loop apex.

**2 Critical magnetic decay index for the flux-rope with fixed endpoints**

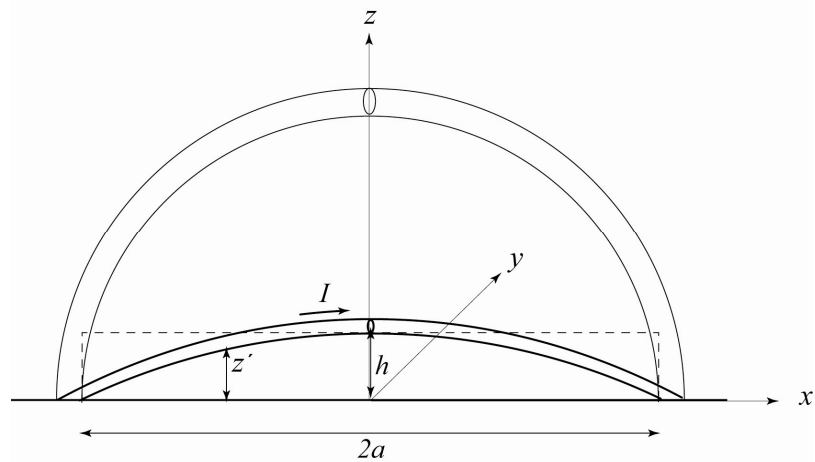
**Figure 1.** Schematic drawing of an erupting circular flux rope

As in several other papers, we consider the flux rope as a section of a torus with its ends embedded in the photosphere, which keeps its partly-circular shape during all evolution (Chen 1989; Cargill et al. 1994; Isenberg and Forbes 2007; Olmedo and Zhang, 2010; Filippov 2020). The major radius of the torus $R$ initially is large because the height of the flux-rope apex $h$ is low

$$R = \frac{a^2 + h^2}{2h}, \tag{1}$$

where $a$ is the half-footpoint separation. We neglect the plasma pressure force in comparison with the Lorentz force (low-beta plasma), as well as the gravity force, which is acceptable for the filament equilibrium in the lower corona, and consider only the equilibrium of the flux-rope apex as a balance of different components of the Lorentz force.

The radial self-force per unit length is given by (Shafranov 1966)

$$F_R = \frac{I^2}{c^2 R}\left[\ln\left(\frac{8R}{r}\right) - \frac{3}{2} + \frac{l_i}{2}\right]. \tag{2}$$

where $I$ is the toroidal electric current, $r$ is the minor radius of the torus. Equation (2) takes into account also the equilibrium along the minor radius of the torus $r$ suggesting the force-free magnetic field structure. The internal inductance per unit length is thereafter chosen $l_i = 1$, which corresponds to the linear force-free internal magnetic structure (Lundquist 1951; Lin et al. 1998; Isenberg and Forbes 2007). This force vanishes at low heights, since the axis curvature becomes large, and nearly straight electric current does not influence itself. On the other hand, the low flux rope is under the strong influence of the diamagnetic photosphere creating the upward force usually modelled by acting of the mirror current (Kuperus and Raadu, 1974)

$$F_I = \frac{I^2}{c^2 h}. \tag{3}$$

Both $F_R$ and $F_I$ are directed upwards, while the external constraining poloidal magnetic field $B_e$ acts downwards with the force

$$F_B = \frac{I}{c} B_e. \tag{4}$$

In the models that do not take into account the mirror current, low-laying flux ropes are unstable (Cargill et al. 1994; Olmedo and Zhang, 2010). The stability of the flux rope near the equilibrium position $h_0$ is determined by the sign of the linearized part of the equation of motion

$$M\frac{d^2 h}{dt^2} = F = F_R + F_I + F_B. \tag{5}$$

Following Demoulin and Aulanier (2010) we write the total force directed upwards as

$$F_R + F_I = f_R I^2/c^2, \tag{6}$$

where

$$f_R = \frac{1}{R}\left[\ln\left(\frac{8R}{r}\right) - 1\right] + \frac{1}{h} \tag{7}$$

is a factor dependent mostly on the geometry of the loop. In contrast to the work of Demoulin and Aulanier (2010) considering a semi-circular current channel with free ends, the factor includes the repulsion from the photosphere.

For small perturbations $z$ around the equilibrium ($h = h_0 + z$) we can write

$$M \frac{d^2 z}{dt^2} = \left. \frac{dF}{dh} \right|_{h_0} z, \tag{8}$$

where the condition $F(h_0) = 0$ is taken into account. In calculation of the total derivative $dF(h, I, r)/dh$, we should take into account the dependence of $I$ and $r$ on $h$:

$$\frac{dF}{dh} = \frac{\partial F}{\partial f_R} \frac{df_R}{dh} + \frac{\partial F}{\partial I} \frac{dI}{dh} + \frac{\partial F}{\partial B} \frac{dB}{dh} = \frac{I^2}{c^2} \frac{df_R}{dh} + \left( \frac{2 f_R I}{c^2} + \frac{B}{c} \right) \frac{dI}{dh} + \frac{I}{c} \frac{dB}{dh}. \tag{9}$$

Conservation of the toroidal flux within the flux rope with the linear force-free field determines the dependence of the flux-rope radius $r$ on the current $I$ (Lin et al. 1998):

$$r = r_0 \frac{I_0}{I}, \tag{10}$$

which leads to the dependence of $f_R$ not only on $h$ but also on $I$

$$\frac{df_R}{dh} = \frac{\partial f_R}{\partial h} + \frac{\partial f_R}{\partial R} \frac{dR}{dh} + \frac{\partial f_R}{\partial r} \frac{\partial r}{\partial I} \frac{dI}{dh}. \tag{11}$$

Using equations (1), (7), and (10) the last expression can be rewritten as

$$\frac{df_R}{dh} = -\frac{1}{h^2} - \frac{1}{R^2} \frac{(h^2 - a^2)}{2h^2} \left( \ln \frac{8R}{r} - 2 \right) + \frac{1}{RI} \frac{dI}{dh}. \tag{12}$$

Equation (8) suggests that the equilibrium becomes unstable when $\left. \frac{dF}{dh} \right|_{h_0} > 0$. So, the threshold of the eruptive instability is defined by the equation

$$\left. \frac{dF}{dh} \right|_{h_0} = 0, \tag{13}$$

together with the equilibrium condition

$$F(h_0) = \frac{I^2}{c^2} f_R + \frac{I}{c} B_e = 0. \tag{14}$$

Using the usual definition of the decay index of magnetic field $n$ (Bateman 1978; Filippov and Den 2000, 2001; Kliem and Torok 2006)

$$n = -\frac{\partial \ln B_e}{\partial \ln h} \tag{15}$$

and substituting Equations (9), (12), (14) into (13) we obtain

$$n_c = \frac{h}{f_R} \left( \frac{1}{h^2} + \frac{1}{R^2} \frac{(h^2 - a^2)}{2h^2} \left( \ln \frac{8R}{r} - 2 \right) \right) + \frac{h}{Bc} \left( f + \frac{1}{R} \right) \frac{dI}{dh} = n_R + n_I. \tag{16}$$

We omit subscript '0' keeping in mind that all values are calculated at the equilibrium height $h_0$.

The dependence $I(h)$ can be found from the conservation of the poloidal magnetic flux between the flux rope and the photosphere, $h = 0$, given by

$$\Phi_p = \Phi_I + \Phi_s = \frac{I}{c} L_e + \int_S B_e ds = \text{const}, \tag{17}$$

where S is the area between the photosphere and the flux rope and $L_e$ is the external self-inductance of the flux rope. The self-inductance of a thin circular conductor of the major radius $R$ and the radius of the cross-section $r$ ($R \gg r$) is defined as (Landau and Lifshitz 1984)

$$L_e = \int_{\varphi_0}^{\varphi} \int_{\varphi_0}^{\varphi} \frac{R \cos \varphi'}{2 \sin \frac{\varphi'}{2}} d\varphi' d\varphi'' = 2R\varphi \left( \ln \tan \frac{\varphi}{4} + 2 \cos \frac{\varphi}{2} + \ln \frac{8R}{r} - 2 \right), \tag{18}$$

where

$$\varphi = 2 \arcsin \frac{a}{R}, \text{ if } h < a$$

$$\varphi = 2\pi - 2 \arcsin \frac{a}{R}, \text{ if } h \geq a \tag{19}$$

and

$$\varphi_0 \approx \frac{r}{2R}. \tag{20}$$

The two first terms in Equation (18) arise from the upper limit of the integration over $\varphi$ and are responsible for the influence of rather far elements of the circuit on a given current element. For a long enough loop they are small compared to the influence of the neighbour elements represented by the last terms. For simplicity the first terms can be omitted as it is accepted in a number of papers (Chen 1989; Cargill et al. 1994; Isenberg and Forbes 2007; Olmedo and Zhang, 2010; Filippov 2020). The difference between the full and shortened expressions is shown in Fig. 2.

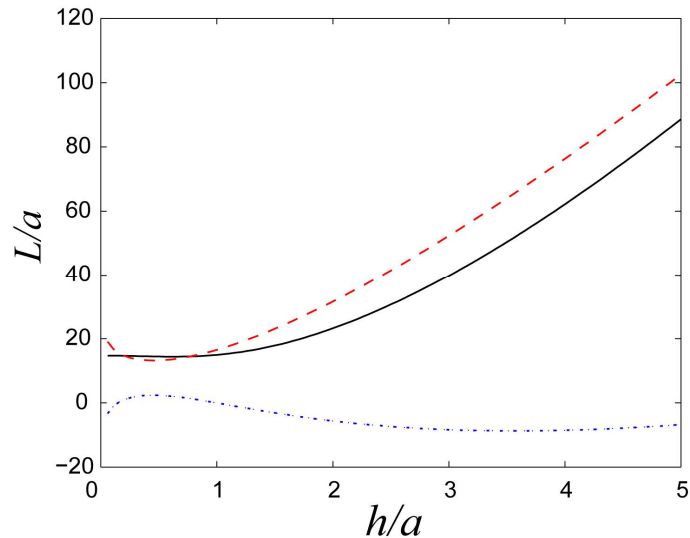

**Figure 2.** Self-inductance of a thin circular loop as a function of the apex height *h*. The solid line represents the full Equation (18), while the dotted (blue) line shows the contribution of the first two terms and the dashed (red) line shows the contribution of the last two terms.

Derivation of Equation (17) with respect to *h* gives

$$\frac{dI}{dh} = -\frac{1}{L_e}\left(I\frac{dL_e}{dh} + c\frac{d\Phi_s}{dh}\right). \tag{21}$$

The total derivative of the self-inductance can be written as

$$\frac{dL_e}{dh} = \left(\frac{\partial L_e}{\partial \varphi}\frac{d\varphi}{dR} + \frac{\partial L_e}{\partial R}\right)\frac{dR}{dh} + \frac{\partial L_e}{\partial r}\frac{\partial r}{\partial I}\frac{dI}{dh}, \tag{22}$$

therefore

$$\frac{dI}{dh} = -I\frac{\left(\frac{\partial L_e}{\partial \varphi}\frac{d\varphi}{dR} + \frac{\partial L_e}{\partial R}\right)\frac{dR}{dh} + c\frac{d\Phi_s}{dh}}{L + I\frac{\partial L_e}{\partial r}\frac{\partial r}{\partial I}}. \tag{23}$$

Finally, substituting Equations (14) and (22) into ((16) and taking the derivatives, we find

$$n_I = \frac{h\left(f + \frac{1}{R}\right)}{L_e + 2\varphi R}\left\{\left[\frac{(h-a)}{abs(h-a)}\left(\frac{L_e}{\varphi} + \varphi\left(\frac{R^2}{a} - 2a\right)\right)\right]\frac{2a}{R\sqrt{R^2 - a^2}} + \frac{L_e}{R} + 2\varphi\right]\frac{(h^2 - a^2)}{2f_R h^2} - \frac{1}{B}\frac{d\Phi_s}{dh}\right\}. \tag{24}$$

For simplicity we assume hereafter the external magnetic field as only a function of height *B(h)* and approximate the value of the flux below the circular contour by the value of the flux through the rectangular contour shown in Fig. 1. Then

$$\frac{d\Phi_s}{dh} = lB, \tag{25}$$

where

$$l = \begin{cases} 2a, & h \leq a \\ 2R, & h > a \end{cases}. \tag{26}$$

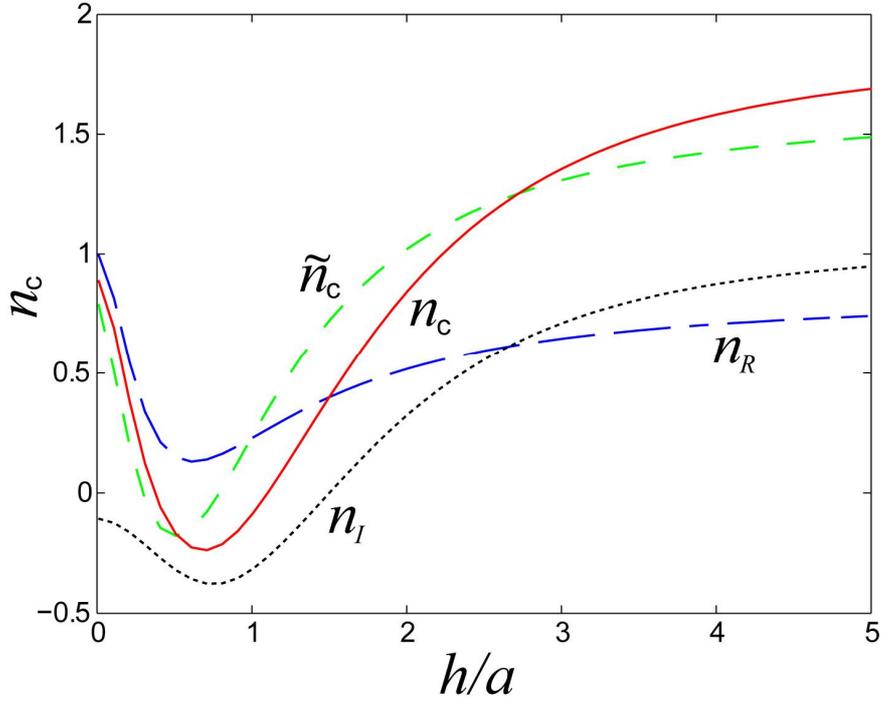

**Figure 3.** Total critical decay index $n_c$ as a function of height (red solid line) and contributions to the total decay index from the geometrical factor of the upward force $n_R$ (blue dashed line), from the current-dependent term $n_I$ (black dotted line), for $r_0 = 0.1a$. Green dashed line shows the total decay index with the expression for the self-inductance $L_e$ without two first terms in the Equation (18).

Figure 3 shows the behaviour of the critical decay index and its component as a function of the apex height in the units of the endpoints half-separation $a$. In contrast to monotonic dependence of the critical index for flux ropes with free ends (Demoulin and Aulanier 2010), the index at first decreases with height from a value close to unity, reaches the minimum lower than zero ($n \approx$ -- 0.25 at $h \approx 0.8a$, then increases exceeding the value of $n_c \approx 1.5$. Low heights correspond to a large major radius and small curvature of the flux rope axis. It is not surprising that the critical index is close to unity, which is typical for a straight current channel. On the other hand, large heights correspond to a nearly fully circular current channel with the typical critical decay index $n_c \approx 1.5$. The low value of the critical index should be inhered to rather short filaments with the conspicuous initial curvature.

**4 Discussion and conclusions**

Coronal magnetic field generated by electric currents flowing below the photosphere plays a significant role in equilibrium of a magnetic flux rope containing a filament in the corona and development of instability leading to an eruption. It was found in theoretical models that a straight current channel becomes unstable if the external magnetic field decreases faster than $h^{-1}$ (van Tend and Kuperus 1978; Molodenskii and Filippov 1987), while a current ring becomes unstable in the field decreasing faster than $h^{-3/2}$ (Osovets 1961; Bateman 1978; Kliem and Torok 2006), or the decay index is equal to 1 or 3/2, respectively. These two idealised cases are far from real geometry of filaments on the Sun. Usually filaments have a smooth loop-like shape with endpoints connected to the photosphere. Models of loop-like flux ropes were suggested by many researchers with different simplifications and approximations. In some of them the external field is not taken into account, and only internal instabilities are considered (Chen 1989; Vrsnak 1990). In the model of Titov and Demoulin (1999) endpoints touch the photosphere, however they are not 'frozen-in' and can slide along the surface so that the height of the apex and major radius of the torus increase synchronously. Cargill et al. (1994) and Olmedo and Zhang (2010) studied a partial current-carrying torus loop anchored to the photosphere. They did not take into account the diamagnetic influence of the photosphere, therefore low-lying loops with a great radius of curvature experience only faint upward force and are unstable. Most rigorous approach is in the model of Isenberg and Forbes (2007), where the line-tying of the magnetic field on the surface is incorporated with great accuracy. Their model exhibits the potential for catastrophic loss of equilibrium as a possible trigger for eruptions, but they did not analyze the dependence of the critical decay index on the parameters of the model.

We used similar but more simplified model than Isenberg and Forbes (2007). The total hoop-force is considered as the sum of the Lorentz forces arising from the action of neighbouring current elements and the influence of the dense photosphere, modelled by a mirror current. This model was used in an analysis of failed filament eruptions (Filippov 2020), where the main attention was paid for possibility of an existence of the second equilibrium point at a greater height after catastrophic loss of equilibrium at a lower height. In this paper we analyze conditions of this loss of equilibrium.

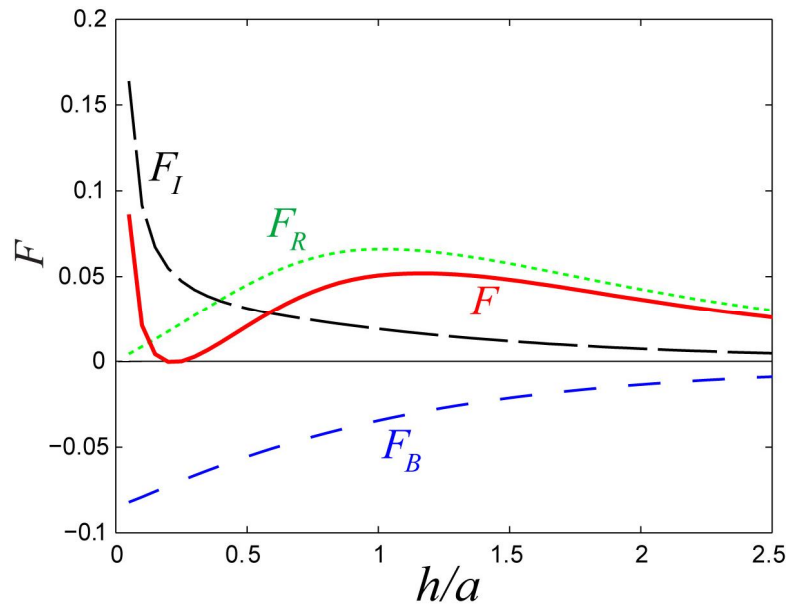

**Figure 4.** Total force $F$ (in dimensionless units) as a function of height for the external field generated by a linear dipole at a depth of $a$ below the photosphere and contributions to the total force from the radial self-force $F_R$, (green dotted line), the repulsion from the photosphere $F_I$ (black dashed line), and the constraining force $F_B$ (blue dashed line).

The shape of the curve describing the dependence of the total force on height is not strongly influenced by the particular distribution of the external magnetic field. Figure 4 shows the total force (5) as a function of height for the external field generated by a linear dipole at a depth of $a$ below the photosphere. The value of the electric current is chosen so that there is a point of catastrophic loss of equilibrium: the curve has the extremum (minimum) point touching the horizontal axis $F = 0$ at $h \approx 0.2a$, where $n \approx 0.35$ (as it is seen in Fig. 3). The shape of the curve is similar to that shown in Fig. 6 in the paper of Isenberg and Forbes (2007), except their extremum point lies below the axis of abscissa allowing two equilibrium points to exist: the lower stable point and the higher unstable one.

While in most models the critical decay index has a value of unity or more, in our model it is much below unity in the large range of heights. Very low values of the critical decay index (even negative) exist due to the changing of the geometrical shape of a flux rope. It transforms from nearly straight to partly-circular with the radius at first decreasing and later increasing. In the models of loop-like flux ropes with free endpoints, the major radius of the torus increases synchronously with the apex height that leads to the critical decay index value greater than one. Anchoring the endpoints results in non-monotonic changing of the loop radius and the complicated dependence of the critical decay index on height. The model suggests very low values (even below zero) of the index when the height of the apex is approximately equal to the half-separation of the endpoints and the shape of the flux rope axis is half-round.

One cause of small values of the critical decay index, when the shape of the current channel is nearly semicircular, is a rapid increase of $F_R$ with height due to the rapid decrease of the curvature radius $R$. The dependence of the critical index on the shape of the channel is demonstrated by the curve $n_R$ in Fig. 3 having a dip below $h/a = 1$. This is the intrinsic property of the model of a partial current-carrying torus loop with the endpoints anchored to the photosphere. In the absence of the mirror current creating the force $F_I$ (as was considered by Cargill et al. 1994; Olmedo and Zhang 2010), stable equilibrium of lower lying loops becomes unrealistic because it needs rapidly increasing with height external magnetic field with a high negative value of the decay index. The presence of the mirror current provides the possibility of the stable equilibrium for a low lying loop in the external field with the decay index of the order of unity. Nevertheless, there is a dip in the critical decay index vertical profile shown in Fig. 3. Therefore, the nearly semicircular shape of the flux-rope axis is the most unfavourable for stable equilibrium.

However, this is strictly valid in the model with the perfectly circular shape of the flux-rope axis. In the real solar atmosphere, the shape of the flux-rope axis in perturbations near an equilibrium position can be not a fragment of a circle. The shape of the flux rope can change not supporting a perfect circular shape as it is assumed in the model. The upper part of the flux rope can keep a more flat shape and rise without significant changing of curvature. In this case, the decrease of the critical index below unity will be less perceptible. For the circular channel with free ends, the critical index increases monotonically with height approximately from 1 to 1.5 (Fig 3d in Demoulin and Aulanier 2010).

The other cause for the low values of the critical index near $h/a \approx 1$ and the presence of the dip in the profile $n_c(h)$ is the specific dependence of the electric current value on height near the photosphere according to the inductance equation (17). While the self-inductance of the flux rope is nearly constant at low heights (Fig. 2), the flux of external field is greater for greater heights of the flux-rope apex. As a result, the electric current increases. It becomes decreasing with height when the self-inductance starts to rise due to the growth of the curvature radius of the flux-rope axis (Fig. 2). Examples of the electric current changes with height in some types of the external field according the induction equation are presented in Figs. 11 and 12 in (Filippov 2020). The behaviour of the part of the critical index dependent on the electric current, $n_I$, is shown in Fig. 3.

It should also be noted that our model does not take into account a gravity force. If the mass of the filament is large enough for the gravity force to be comparable with the Lorentz force, values of the critical index become greater as was shown by Tsap et. al. (2019). Low values of the decay index at the height from which prominences start eruption were reported by Lee et al. (2016), Vasantharaju et al. (2019), Rees-Crockford et al. (2020), Filippov (2020).

Analysis of the 3D magnetic field distribution around solar filaments is important for estimations of their stability and possibility of eruptions. As reliable measurements of the magnetic field are available at present time only in the photosphere, development of extrapolation techniques is needed for reconstruction of the magnetic-field structure in the corona. Different models of magnetic flux-rope equilibrium suggest different criteria for the instability threshold. New observations of eruptive processes and precise calculations are need for deeper understanding of the initiation of the eruptions.

## 4 DATA AVAILABILITY

There are no new data associated with this article.